# Dynamic Prioritization of Emergency Vehicles For Self-Organizing Traffic using VTL+EV *

Subash Humagain and Roopak Sinha
*IT and Software Engineering department*
*Auckland University of Technology*
*Auckland, New Zealand*
subash.humagain@aut.ac.nz, roopak.sinha@aut.ac.nz
ORCID: 0000-0002-1652-930X, 0000-0001-9486-7833

**Abstract**

*Cooperative vehicular technology in recent times has aided in realizing some state-of-art technologies like autonomous driving. Effective and efficient prioritization of emergency vehicles (EVs) using cooperative vehicular technology can undoubtedly aid in saving property and lives. Contemporary EV prioritization, called preemption, is highly dependent on existing traffic infrastructure. Accessing crucial decision parameters for preemption like speed, position and acceleration data in real-time is almost impossible in current systems. The connected vehicle can provide such data in real-time, which makes EV preemption more responsive and effective. Also, autonomous vehicles can help in optimizing the timing in traffic phases and minimize human-related loss like higher headway times and inconsistent inter-vehicle spacing when following each other. In this paper, we introduce self-coordinating a decentralized traffic control system termed as Virtual Traffic Light plus for Emergency Vehicle (VTL+EV) to prioritize EVs in an intersection. The proposed system can expedite EVs movement through intersections and impose minimal waiting time for ordinary vehicles. The VTL+EV algorithm also can improve overall throughput making an intersection more efficient.*

## 1. INTRODUCTION

Cooperative vehicle technology has helped advanced applications to evolve in recent times. Advancements in wireless communication techniques that use dedicated short-range communications (DSRC) for vehicular ad-hoc networks (VANET) have already found application in the real-world. Multiple applications like electronic brake lights (allowing drivers or autonomous vehicles react to obstructions by enforcing braking), platooning (following a leader vehicle within inches utilizing real-time exchange of acceleration and speeding information), emergency response services (allowing emergency vehicles to respond on time) and add-on services (advertising for restaurants, petrol stations, etc., to the driver) are already being used in the automotive industry using the IEEE 802.11p communication protocol [1]. These applications eventually increase road safety and elevate overall traffic management.

Traffic management and road safety are the principal problems traffic engineers and planners struggle to solve. According to the World Health Organization's Global status report in 2018, 2.34 million road users lost their lives in 2016. It also states that road injury has emerged as the eighth major cause of death for people of all ages and the number 1 cause of death among children and adults aged between 5-29 years [2]. Studies show that more than 90% of road accidents are due to human errors [3]. Intersections are more prone to accidents because of lack of surveillance (44.1%), the wrong assumption of others' movement (8.4%), movement in obstructed view (7.8%), not following the priority rules (6.8%), distractions within the vehicle (5.7%) and misinterpretation of inter-vehicle gaps (5.5%) [4]. Existing technology and measures to reduce incidents at intersections seem insufficient. In case of accidents and emergencies, the traffic control system must aid EVs to expedite movement. In current practice, it is achieved by prioritizing EVs movement, which is termed as Emergency Vehicle Preemption.

Contemporary preemption techniques are inefficient and infrastructure dependent. Different preemption systems like Tramsmax, OPTICOM, GERTRUDE, and FAST are currently being used in multiple cities of UK, USA, Australia, Canada and Japan [5]. These solutions provide empirical evidence of reducing the overall response time of EVs, but all of these solutions can be realized only with the installation of additional devices. The presence of EV is detected from different sensors like localized radio, line of sight or acoustic sensors and GPS installed within vehicles or near the intersections. The traffic is usually controlled from centralized traffic control centres using information sent by these sensors and priority is assigned to EV by altering the traffic signals. Unlike the centralized approach, some systems like EMTRAC [6] use traffic intersections to decide locally on preemption. Decentralized systems can be operated without any centralized backbone network connecting all the traffic intersections. However, this kind of system has two obvious disadvantages:

- Individual intersections have higher operation, maintenance and installation costs.
- No coordination among the intersections results in sub-optimal preemption.

Innovative state-of-art technologies like cooperative vehicular technologies and autonomous driving can mitigate limitations of current preemption techniques elevating efficiency and road safety. Implementations of VANET help in realizing infrastructure-free self coordinating traffic control. An ad-hoc wireless network generated form vehicular movement communicating with each other or infrastructure is termed as VANET. Vehicles and infrastructure use IEEE 802.11p communication protocol to communicate with each other using DSRC devices [7]. Vehicles can use information beacons with speed, position and direction data transmitted within VANET to coordinate their movement in a traffic intersection. Decisions regarding stopping, acceleration, and turning are adopted by vehicles if they are autonomous or informed to the drivers using inbuilt display devices installed within vehicles. Such self-coordinating traffic control system does not rely on costly infrastructures like traffic lights and any backbone network infrastructures and is termed as virtual Traffic Lights (VTL). Contemporary VTL systems propose a traffic intersection control system that improves the overall efficiency in terms of waiting time and throughput.

The VTL concept has been realized in real-world traffic conditions. The concept of self-organized traffic control termed as VTL was first introduced in [8] where a leader vehicle was elected which acts as a temporary traffic controller and generates traffic light signals using the information available via VANET and sends traffic light information to the drivers or in-vehicle display units. An extension to provide priority to EVs once detected near intersection using VTL was implemented in [9]. Through simulation results they claimed that the travel time of EVs was reduced significantly and impact over general traffic was marginal. A VTL system, software and apparatus required for coordinating traffic approaching a conflicting zone was designed and patented. The system developed a dynamic traffic plan and sent to the in-vehicle display unit, which controls the vehicle to follow the plan [10]. The system described in [10] was implemented for the real-world trials in the street of Pittsburgh and was proved that the developed system was capable of coordinating traffic at intersections and reduce the commute time [11].

VTL is gaining popularity among researchers and innovators. Shi et al. introduced a concept where vehicle express their will of moving forward, and the leader provides the way according to the score of will that depends on the dynamic traffic condition of the intersection [12]. Instead of using VANET, the conceptual model of implementing VTL using mobile communication and cloud server was introduced in [13]. A distributed algorithm that uses both broadcast signals and unicast messages for assigning priority to vehicles that struggle to resolve movement conflicts while approaching an intersection were proposed in [14] and [15]. A VTL framework that can work for both DSRC enabled modern vehicles and normal cyclist and pedestrians was proposed in [16]. The traffic control information to these users like cyclist and pedestrian was transmitted to their smartphones using Bluetooth devices. In addition, [17] developed a graphical user interface that projects VTL sequences on the windscreen of the vehicle using head-up displays and [18] identified the optimal area within the vehicle to place the VTL. Eventually, Sinha et al. identified two issues for the adoption of the VTL in industry, mainly functional safety analysis and migration from non-equipped vehicles to VTL. The solution for the first issue was proposed using a model-driven engineering approach. The solution to the second problem was proposed as implementing VTL+ that uses additional vehicle-to-infrastructure communication from existing infrastructure [19].

Very less has been studied about prioritizing EVs using VTL. Most of current VTL system elects a leader within an intersection that process the speed and distance information available from multiple vehicles and sends the scheduling instructions to others. In case of accidents, VTLs should aid on prioritizing EVs movement by instructing other vehicles around the accident scene. So far to our knowledge, [9] is the only study that has implemented a self-organized traffic control system that manages the priority of EVs. They implemented a conventional static traffic control system with traffic lights to a VTL decentralized to a single leader vehicle stopped at an intersection waiting for the green phase. They utilized the dead periods (unwanted green signal to complete the phase time even if there are no vehicle from that approach) to show a significant increase in traffic throughput. In their approach, an EV approaching an intersection broadcasts priority request message, the leader replies with acknowledgement message, halts regular signal operation and assigns priority by granting green phase to EV. Once EV passes the intersection, it sends a clear message, and regular traffic signal operation is resumed.

In this study, we propose a novel self organizing traffic control system. We consider the vehicles can manoeuvre autonomous driving. Unlike current VTL (replication of traditional traffic lights controlled by a lead vehicle at an intersection), the Virtual traffic light plus for EV (VTL+EV) algorithm we propose completely eliminates the everlasting traffic signalling concept of optimized cycle time and phase duration. Vehicles approaching an intersection calculate the exact speed and duration to pass the intersection and adapt accordingly. Vehicles from any direction should not wait until they get a green phase to pass. This eliminates unnecessary waiting at dead periods and makes the entire system adaptive. Since vehicles are autonomous, they can maintain a minimal distance when they follow other vehicles and the time allowed for a human driver to respond to the change in traffic phase (headway time) can be neglected. The closely following group of vehicles are divided into platoons and each platoon gets a reserved time to pass the intersection. Since there is no phase change, any platoon from any direction can pass the intersection minimizing all the delays experienced in the traditional traffic control system. The speed maintained by each platoon is dependent on the platoon crossing the intersection. If there is any platoon of vehicles crossing the intersection, the next platoon to cross maintains such a speed so that it does not collide with the platoon crossing the intersection. This ascertains that the intersection is thoroughly utilized and does not allow unnecessary waiting for vehicles when it is free due to inappropriate phase cycles.

VTL+EV dynamically prioritizes EVs approaching an intersection. Whenever an EV approaches the intersection, it transmits information beacon that contains its current position, speed and ID (code to identify it as EV). This halts the current operation of the traffic controller and calculates reserve time for EV to cross the intersection from the current location and allows traffic from this direction only. Once EV crosses the intersection, it resumes its regular operation. In this approach of prioritizing EVs (self-organized traffic control) vehicles from any direction Zip while crossing the intersection. This minimizes all the delays that exist in contemporary systems. The effect of preemption on other vehicles is negligible.

The proposed system shows promising results. We considered well known four-legged traffic intersection to perform multiple experiments using microscopic traffic simulation engine called SUMO [20]. We compared average waiting time of EVs and normal vehicles in VTL+EV system with self-organized Traffic control system called as VTL-PIC to manage the priority of EVs [9] and enhanced traffic light scheduling algorithm (ETLSA) that assigns adaptive green phase to traffic according to the number of traffic flow [21]. We also compared the overall throughput and queue-length of intersection implementing each of above-mentioned EV's preemption system. VTL+EV algorithm outperformed both VTL-PIC and ETLSA preemption systems. The average waiting time was considerably less, throughput was significantly high, and average queue-length was minimal.

## 2. SYSTEM MODEL

A four-legged intersection is the most commonly used model in traffic engineering for studying the performance as it can be scaled to simpler or more complex intersections. Whenever any vehicle (both autonomous and manual driven) approaches an intersection, it examines the traffic phases. If the traffic phase is green, it crosses the intersection else manoeuvres deceleration or stops altogether. In case of manual vehicles, the driver uses senses to identify the traffic phases whereas autonomous vehicles combine signals from multiple sensors like lidar, sonar, radar, odometer, GPS and inertial measurement units to identify appropriate driving actions [22]. In our approach, we consider that every intersection comprises of an Intersection Traffic Controller (ITC) that communicates with incoming platoons of autonomous vehicles sending instructions regarding stopping, acceleration and turning. The same system can be deployed for manual vehicles enabled with DSRC devices. However, it will result in less efficiency because of involved human factors like increased headway time and inconsistent platoon size.

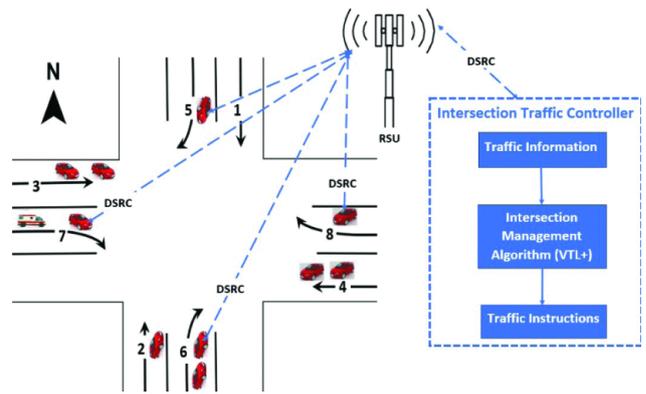

Fig. 1. Intersection Traffic Control System

In this study, we assume that all the vehicles are equipped with DSRC communication devices to communicate position, speed, and vehicle type data collected from inbuilt vehicle sensors like GPS. Every ITC has a defined range of area. A roadside unit (RSU) collects traffic information and supplies it to the ITC. The communication between vehicles and RSU is carried out in the 5.9GHz spectrum dedicated for DSRC using IEEE 802.11p. These data are used by platooning algorithm and traffic control algorithm to generate traffic instructions, which are transmitted back to vehicles to adhere to. Fig. 1 reflects the basic system operation. Whenever an EV shows up within the range of the ITC, it notifies the RSU. The RSU distinguishes EVs from other vehicles using its vehicle type parameter which is set to Emergency. Once detected, the traffic control algorithm calculates the reserved time required for this particular EV to cross the intersection. Updated speed, acceleration and stopping instructions are sent to the EV and all other vehicles. Once the EV crosses the intersection, it sends an acknowledgement message to the ITC to resume normal operation.

### A. Platooning

Platooning increases road capacity by eliminating constant stop and go of individual vehicles. Modern traffic management systems have leveraged a lot from inter-vehicle communication technologies. The advent of autonomous vehicles is one of the promising examples to cite. Inter vehicle communication has become very reliable; that is why autonomous vehicles can follow each other within a distance of inches forming platoons. Autonomous vehicles abolish time loss due to human factors like slow reaction time, diverse driving behaviour and possibilities of attention lapses. Platooning, in addition, increases traffic efficiency [23]. To leverage the advantages mentioned above, instead of implementing self coordinating traffic for individual vehicles, we have divided vehicles approaching an intersection into platoons.

The algorithm to construct platoons is depicted in Algorithm. 1. The area within the range of ITC is divided into different controlled lanes. Controlled lanes are gradually added with vehicles one at each time step. The first vehicle, so added, converts into a single sized platoon. If two platoons in a controlled lane are within a certain distance and share the same route, they are eligible for merging. In that case, two existing platoons are disbanded, and a new platoon of size two is created. This condition is checked until the size of platoon increases to a maximum

**Algorithm 1** Platooning Algorithm

**Input:** startingVehicles, active, color, currentSpeed, disbandReason, eligibleForMerging, lane, lanePosition, controlledLanes, and targetSpeed
**Output:** Platoons
1: **for** each ControlledLanes $l$ **do**
2:   addVehicle (vehicle);
3:   mergePlatoon($P_i$):
4:     **if** ($l.P_i.VehPathsConverge(l.P_i.getAllVehicles())$ && $l.P_i.getLane() == l.P_i.getLane()$ ) **then**
5:       $l.P_i.disband()$;
6:       $l.P_i.disbandReason = Merged$;
7:       **for** vehicle in $l.P_i.getAllVehicles()$ **do**
8:         $l.P_i.addVehicle(vehicle)$;
9:     **else**
10:       $l.P_i.eligibleForMerging = False$
11:   setTargetSpeed(speed);
12:   setGap(gap);
13:   updateSpeed(speed, inclLeadingVeh=True);

**Algorithm 2** VTL+EV Algorithm

1: **for** this Controller $c$ **do**
2:   addPlatoon ($P_i$);
3:   calculatePlatoonReservedTime ($P_i$);
4:   removeUncontrolledPlatoons ( );
5:   addAllControlledPlatoons ( );
6:   setZipOrderForController ($P_i$ ):
7:     **if** $((c.P_i.getVehicletype()) == Emergency)$ **then** process $P_i$;
8:     **else** distSort(elem):
9:       return elem.getLanePositionFromFront()
10:   createZipPlatoons ( );
11:   getlaneposition ($P_i$);
12:   setNewSpeed ($P_i$, reservedtime);
13:   removePlatoon ($P_i$);
14:   update ( );

allowable size. Literature suggests that platoon length of a maximum of 35 meters does not diminish overall performance [23]. Therefore, considering a single-vehicle size of 3 meters, we have defined platoon size to 12 vehicles. Every platoon has a leader. Other vehicles within the platoon follow leader's behaviour within the set gap in terms of speed and acceleration. Once a platoon gets outside the ITC zone, vehicles continue being within the same platoon if they still satisfy the above-mentioned condition else they disband themselves.

**B. Proposed VTL+EV Algorithm**

ITC sets instructions regarding speed required for each platoon to adhere while crossing an intersection using the VTL+EV algorithm. An area is defined for the intersection controller. We add platoons from all lanes and calculate the reserved time required for the platoon to cross the intersection. If this platoon is the first one to post the reservation, we calculate distance from the stop line to the platoon's current position and add it with the platoon's length which is done as shown in Fig. 2. If this platoon contains EV we allow this platoon to pass else we process platoons in ascending order of its distance from the stop line. The VTL+EV algorithm is illustrated in Algorithm 2. Every platoon within the influence of VTL+EV algorithm is allocated with the time calculated from reserved time to pass the intersection.

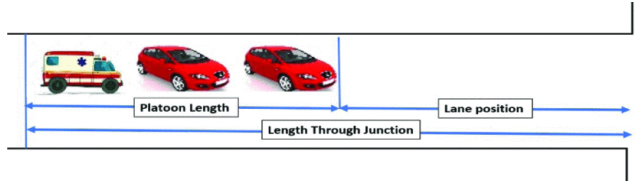

Fig. 2. Calculation of reserved time

## 3. IMPLEMENTATION

We compare the performance of VTL+EV algorithm with self-organized Traffic control system called as VTL-PIC and enhanced traffic light scheduling algorithm (ETLSA) using well known microscopic traffic simulator SUMO. The key performance parameter we compared was the average waiting time measured in seconds for both EVs and general traffic. The efficiency of the intersection controller was measured in terms of throughput expressed in passenger car unit per lane per hour (pcu/ln/sec).

SUMO is an open-source continuous microscopic traffic simulator. SUMO gathers information on aspects like networks, routes, trips, and additional sensor devices from its components in advance and runs a simulation until completion. We cannot alter any parameters when the simulation is on. We used an additional tool implemented using Python called Traffic Control Interface (TRACI). TRACI gives access to the live road traffic simulation, allows to capture the required values of simulation and changes their behaviour while the simulation is live. TRACI and SUMO communicate during the simulation using virtual ports (TCP sockets) following the TCP/IP protocol. SUMO behaves as a server that provides services to the client TRACI whenever it sends any service request. As in client-server, architecture SUMO can handle multiple TRACI request at the same time. TRACI also helps to connect different network simulators like NS-3 and OMNET++ with SUMO.

Inter-vehicle communication is essential for the implementation of VTL+EV. A wireless network VANET created because of vehicle communicating with other vehicles and infrastructure is the backbone of the VTL+EV algorithm. This can be realized using an open-source framework Veins that connects OMNET++ and SUMO using TRACI. Using TRACI inbuilt functions, we can directly access traffic network information like lanes, vehicle lane position, speed, acceleration, vehicle ID and vehicle type. We can use this information to execute both platooning and VTL+EV algorithms. In this study, we study if VTL+EV enhances the identified traffic parameters while assuming robust and dependable VANET communications. Since we are not interested in the analysis of wireless communication performance, we do not implement traffic simulation through Veins (an open source framework based on SUMO and OMNET++ for implementation of different models of inter-vehicle communication) and instead use TRACI to access this information. Fig. 3 visualizes the simulation of VTL+EV algorithm using SUMO and TRACI.

We choose traffic parameters during our simulations to replicate real-world traffic scenarios. During the entire course of simulation we first kept on increasing the traffic volume from low (400 pcu/ln/hr), to medium (800 pcu/ln/hr), and then high (1600 pcu/ln/hr) and subsequently decreased it back to medium and then low. Vehicles are generated randomly using inbuilt python command randomTrips.py and the result follows Binomial distribution approximating to Poisson distribution for small

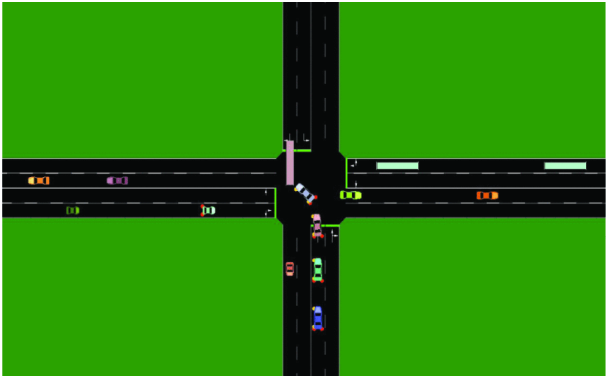

Fig. 3. Implementation of VTL+EV algorithm in SUMO

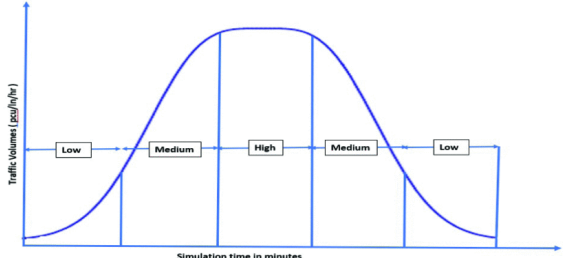

Fig. 4. Traffic volume in Poisson distribution

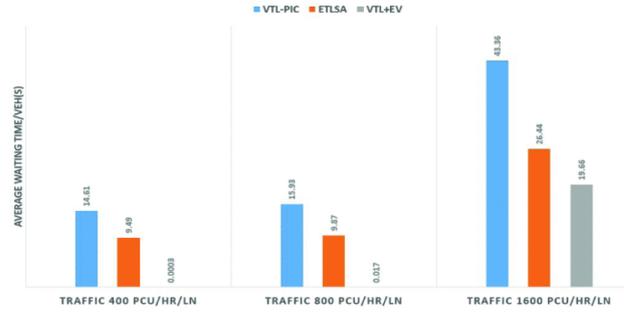

Fig. 5. Average waiting time of normal traffic

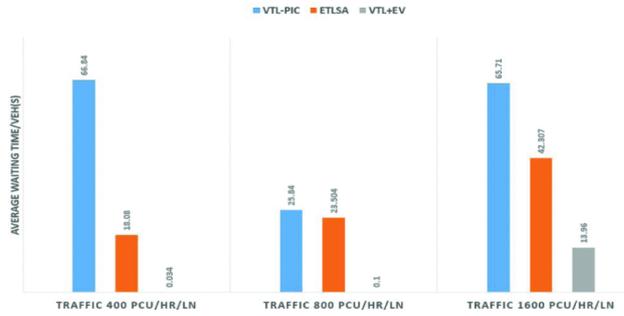

Fig. 6. Average waiting time of EVs

probabilities as expressed in Fig. 4. We have considered 1% of total vehicles generated as EVs.

## 4. PERFORMANCE EVALUATION

VTL being a substitute for physical traffic lights has the advantage of being more economical, but to compare its performance with a simplistic system would be implausible as there are already more advanced adaptive traffic control systems. Therefore to justify that the results produced by VTL+EV are promising we compare its performance with the following state-of-art technologies implemented earlier:

- VTL-PIC: Basic VTL implementation with the use of DSRC technology to detect the presence and absence of EVs at an intersection and assign priority to EVs [9].
- ETLSA: An adaptive traffic scheduling algorithm that dynamically adjusts green phases of traffic signals based on real-time traffic distribution and allows EVs to pass smoothly by coordinating traffic signals using VANET [21].

We conduct multiple experiments with different traffic penetration rates to evaluate below-listed performance parameters for a four-legged traffic intersection:

- *Average Waiting Time* of all Non-EVs.
- *Average Waiting Time* of all EVs.
- *Average Queue length* of intersection under study.
- *Overall Throughput of intersection under study.*

Fig. 5 dissipates the overall waiting time of non-EVs for VTL-PIC, ETLSA, and VTL+EV algorithms. Since VTL-PIC is a virtual representation of a traditional preemption system, such a system incurs a very high waiting time for non-EVs. ETLSA, being the most recent adaptive traffic control system that prioritizes EVs, claimed to outperform existing VANET enabled traffic control system. The increased waiting time resulting from static preemption has been considerably reduced in ETLSA. However, VTL+EV outperforms both of these algorithms in terms of waiting time as VTL+EV is more adaptive and eliminates the loss that arises from multiple traffic phase change and human-related factors. For low and medium traffic volumes, waiting time of VTL+EV is almost zero as autonomous vehicles prefer maintaining the required speed to cross the intersection than to stop. We also compared the average waiting time of EVs in all of these systems. VTL-PIC still used the traditional approach of continuous cycle time. The average waiting time for EVs in this system is highest because the controller must complete the current traffic phase before changing it to green for providing EVs uninterrupted passage. VTL+EV still outperforms both VTL-PIC and ETLSA in terms of average waiting time for EVs, as illustrated in Fig. 6.

Queue lengths and throughput represent elementary performance parameter for quantifying the capacity of any traffic controller. Queue length is measured as the difference in the number of vehicles approaching and leaving an intersection. We compared the average queue length of VTL+EV with VTL-PIC and ETLSA. The simulation results show that the queue length for VTL+EV under all traffic penetration rates is considerably less than VTL-PIC and ETLSA as displayed in Fig. 7. Similarly, throughput computes the traffic volume that an intersection can process within a specific time. Our experimental results show that VTL+EV has the highest throughput compared to the remaining two algorithms under comparison, as pictured in Fig. 8.

## 5. CONCLUSION AND FUTURE WORK

We implemented a decentralized self coordinating traffic system to prioritize emergency vehicle movement through an isolated traffic intersection. The proposed Virtual traffic lights plus for emergency vehicles (VTL+EV) algorithm for intersection control eliminates the loss generated from dead periods in a traffic light cycle time and human-related factors like increased headway time and inconsistent inter-vehicle spacing. We conducted comprehensive experiments and results showed that VTL+EV has the evident advantage of reduced waiting time for regular traffic as well as emergency vehicles (EVs). The overall throughput of VTL+EV implemented traffic intersection are higher and experiences fewer queue lengths. In future, we aim to realize VTL+EV algorithm by implementing it in larger calibrated city maps.

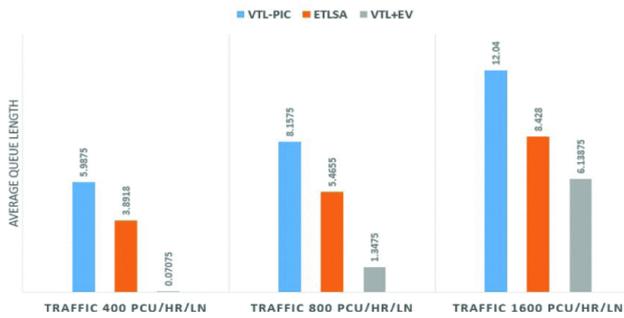

Fig. 7. Average queue length

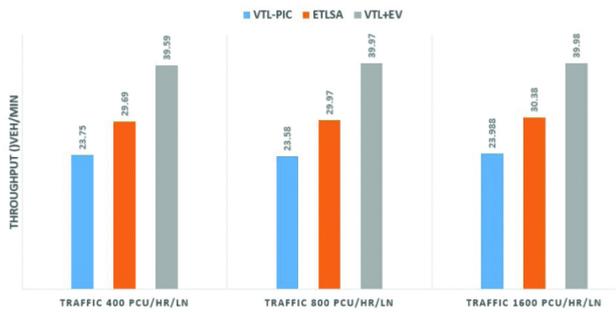

Fig. 8. Average throughput